\documentclass[12pt,preprint]{aastex}

\def\ifundefined#1{\expandafter\ifx\csname#1\endcsname\relax}

\newif\ifpdf
\ifx\pdfoutput\undefined
   \pdffalse      
\else
   \pdfoutput=1   
   \pdftrue
\fi

\def\la{\mathrel{\hbox{\rlap{\hbox{\lower4pt\hbox{$\sim$}}}\hbox{$<$}}}}
\def\ga{\mathrel{\hbox{\rlap{\hbox{\lower4pt\hbox{$\sim$}}}\hbox{$>$}}}}

\newcommand{\be}{\begin{eqnarray}}
\newcommand{\ee}{\end{eqnarray}}

\ifundefined{ensuremath}\def\ensuremath#1{\relax\ifmmode{#1}}
\else${#1}$\fi\else\relax\fi
\ifundefined{nuc}\def\nuc#1#2{\relax\ifmmode{}^{#1}{\protect\text{#2}}
\else${}^{#1}$#2\fi}\else\relax\fi

\newcommand{\etal}{et al.}

\newcommand{\kmps}{km~s$^{-1}$}

\newcommand{\nni}{\nuc{56}{Ni}}
\newcommand{\xni}{\ensuremath{\mbox{X}_{\mbox{Ni}}}}
\def\ang{\hbox{\AA}}
\def\Tmod{\ensuremath{T_{\mbox{model}}}}
\def\Teff{\ensuremath{T_{\mbox{model}}}}

\def\tstd{\ensuremath{\tau_{\mbox{std}}}}

\newcommand{\vno}{\ensuremath{v_0}}

\newcommand{\phx}{\texttt{PHOENIX}}

\received{} 
\accepted{} 
\journalid{}{} 
\articleid{}{} 
\shortauthors{Baron, E. et~al.}
\shorttitle{Metallicity and Mixing in SN 1993W}

\begin{document}

\title{Determination of Primordial Metallicity and Mixing in the Type
IIP Supernova 1993W\footnote{Partially based on observations performed
at ESO La Silla}}

\author{E. Baron\altaffilmark{2}\email{baron@nhn.ou.edu} 
Peter~E. Nugent\altaffilmark{3}\email{penugent@lbl.gov}
David Branch\altaffilmark{2}\email{branch@nhn.ou.edu} 
Peter H.~Hauschildt\altaffilmark{4,5}\email{phauschildt@hs.uni-hamburg.de}
M.~Turatto\altaffilmark{6}\email{turatto@pd.astro.it}
and\\
E.~Cappellaro\altaffilmark{7}\email{cappellaro@na.astro.it}}

\altaffiltext{2}{Department of Physics and Astronomy, University of
Oklahoma, Norman, OK 73019-0260}

\altaffiltext{3}{Lawrence Berkeley National Laboratory, Berkeley, CA
94720}

\altaffiltext{4}{Department of Physics and Astronomy \& Center for
Simulational Physics, University of Georgia, Athens, GA 30602, USA}

\altaffiltext{5}{Present Address: Hamburger Sternwarte, Gojenbergsweg 112,
21029 Hamburg, Germany}

\altaffiltext{6}{Osservatorio Astronomico di Padova, vicolo
dell'Osservatorio 5, I-35122 Padova, Italy}

\altaffiltext{7}{INAF - Osservatorio Astronomico di Capodimonte,
salita Moiariello 16 80181, Napoli, Italy}

\begin{abstract}
We present the results of a large grid of synthetic spectra and compare
them to early spectroscopic observations of SN~1993W. This supernova
was discovered close to its explosion date and at a recession velocity
of 5400~\kmps\ is located in the Hubble flow. We focus here on two
early spectra that were obtained approximately 5 and 9 days after
explosion.  We parameterize the outer supernova envelope as a power-law
density profile in homologous expansion.  In order to extract
information on the value of the  
parameters a large number of models was required. We show that very
early spectra combined with detailed 
models can provide constraints on the value of the power law index, the
ratio of hydrogen to helium in the surface of the progenitor, the
progenitor metallicity and the amount of radioactive nickel mixed into
the outer envelope of the supernova. The spectral fits reproduce the observed
spectra exceedingly well. The spectral results combined with the early
photometry predict that the explosion date was $4.7 \pm 0.7$~days
before the first spectrum was obtained. The ability to obtain the
metallicity from early spectra make SN~IIP attractive probes of
chemical evolution in the universe and by showing that we have the
ability to pin down the parameters of the progenitor and mixing during
the supernova explosion, it is likely to make SN~IIP useful
cosmological distance indicators which are at the same time 
complementary to SNe~Ia.

\end{abstract}


\section{Introduction} 
Supernova 1987A (SN~1987A) confirmed our basic theoretical
understanding of a Type II supernova (SN~II) as the core collapse of a
massive star, which leaves behind a compact object (neutron star or
black hole) and expels the outer mantle and envelope into the
interstellar medium \citep[see][and references therein]{sn87arev89}. While the explosion mechanism
itself remains a subject of active research, the propagation of the
shock wave through the mantle and envelope is reasonably well
understood. Theoretical models for the light curve do a good job of
reproducing the observations
\citep{blinn87a99a,blinn87a00} and detailed radiation transport
calculations verify that these models reproduce the observed spectra
from the UV to the IR \citep{mitchetal87a01,mitchetal87a02}. Unlike
Type Ia supernovae 
(SNe~Ia) where the actual compositions as a function of velocity are
an important subject of current research \citep{fisher91T99}, the
compositions of the envelopes of red or blue supergiant stars is
primarily hydrogen and helium and thus we show that
the primordial compositions, and the degree of mixing of hydrogen
and \nni, can be determined to high precision by the detailed spectral
modeling of observed SN~II spectra. 

SNe~II have a very large spread in their intrinsic brightness, from the
very dim SN~1987A, to the exceedingly bright SNe~1979C and 1997cy. The
observed spread in intrinsic luminosity is greater than a factor of
500. This is not surprising given the fact that the progenitors span a wide
range of initial stellar masses, possible binary membership, and prior
star formation histories. Clearly, SNe~II do not meet the astronomers
requirement of being a \emph{standard candle}, however we believe
that the fact that our models can determine the stellar compositions,
degree of mixing, 
and kinetic energy of the explosion shows that their atmospheres can
be well understood. Although we do not present
distances in this paper, we believe that these results
increase the attractiveness of SNe~II as cosmological probes and in
the future SNe~II will become complementary with
SNe~Ia as distance indicators. Both the SEAM \citep[see][and
references therein]{b93j4,b94i1,mitchetal87a02} and the EPM
\citep[see][]{baadeepm,skeetal94,hamuyepm01,leonard99em02} methods for
determining distances to SNe~II (and other supernova types for SEAM)
depend on the ability to model the spectral energy distribution (SED) of the
supernova atmosphere accurately. Clearly, quantities like the
composition and degree of mixing play a role in the output SED and
thus the work presented here helps to place both these methods on
firmer footing.

While the average SN~II is several times dimmer than a SN~Ia, current
ground-based searches and proposed future space-based searches for
supernovae will easily detect these objects at cosmologically
interesting distances. In the very deep high-redshift supernova
searches by the Hi-Z Team \citep{riess_scoop98} and the Supernova
Cosmology Project (SCP) \citep{perletal99} several SNe~II have been
discovered even though the searches are geared to finding SNe~Ia. The
SCP has positively identified via spectroscopy 5 SNe~II out to a
redshift $z \approx 0.45$. \citet{hdf_gnp99} detected
SN~1997fg through differencing two epochs of Hubble Deep Field North
$I$-band photometric images. While the supernova was not
spectroscopically classified, the combination of the host galaxy type
(an irregular), its redshift $z = 0.952$, and the supernova's apparent
brightness makes it a strong candidate for a SN~II. A. Clocchiatti
\etal\ (IAUC 7549) found SN~2000fp at $z=0.30$, G. Altavilla \etal\
(IAUC 7762) found SN~2001gg at $z=0.61$, SN~2001gh at $z\approx 0.16$,
and SN~2001gj at $z=0.27$, and G. Altavilla \etal\ found SN~2002co at
$z=0.318$. The proposed 
\emph{SNAP} satellite, a wide-field optical imaging telescope, (see
{\tt http://snap.lbl.gov}) will easily be able to find, follow, and
spectroscopically identify SNe~II beyond redshifts of $z \approx
1$. \emph{NGST}, a large, small field IR telescope, will be able to
detect SNe~II to the initial star formation period in the universe, $z
\ga 5$.

The explosion mechanism is now thought to result in a non-spherically
symmetric shock-wave \citep{lbw70,khok_jet99} and this has
been confirmed by the detection of significant polarization in the
spectra of SNe~II \citep{wang_pol96,leonmd00}. Nevertheless, the
large expansion of the ejecta before the supernova becomes visible
leads to significant ``sphericalization'' \citep{chev84}, and
asphericity effects may only be important at late times, or in SNe~II
that are strong circumstellar interacters (SNe~IIn). These supernovae
are clearly distinguishable observationally from ``normal'' Type II
supernovae.

Calculations \citep{miralda97,DF99,livyoung00} indicate that SNe~II may in
fact be the first stellar objects visible in the universe and hence
they serve as important probes for the star formation rate, the rate
of chemical evolution by measuring their primordial abundances, and
directly for the cosmological parameters.  We
show that the primordial abundances of the progenitor star can be
reliably determined by detailed synthetic spectral modeling.

\section{Calculations}

We have chosen to model SN~1993W in detail because there are several
high signal to noise spectra very near the time of explosion as well
as at later times and because we found that this object is clearly a
low metallicity supernova although both the photometry (see
Table~\ref{tab:obphoto}) proves that it is indeed an SN~IIP, closely
resembling the light curve of SN~1969L and SN~1988A.

The calculations were performed using the multi-purpose stellar
atmospheres program \phx~{version \tt 11.7}
\citep{hbjcam99,bhpar298,hbapara97,phhnovetal97,phhnovfe296}.
\phx\ solves the radiative transfer equation along characteristic rays
in spherical symmetry including all special relativistic effects.  The
non-LTE (NLTE) rate equations for many ionization states are solved
including the effects of ionization due to non-thermal electrons from
the $\gamma$-rays produced by the  radiative decay of $^{56}$Ni, which
is produced in the
supernova explosion.  The atoms and ions calculated in NLTE are: H~I,
He~I--II, C~I-III, N~I-III, O~I-III Na~I-III, Mg~II, Ca~II, Si~I--III,
S~I--III, Fe~I--III, Ni~I-III, and Co~I-III. These are all the
elements whose features make important contributions to the observed
spectral features in SNe~II.

Each model atom includes primary NLTE transitions, which are used to
calculate the level populations and opacity, and weaker secondary LTE
transitions which are are included in the opacity and implicitly
affect the rate equations via their effect on the solution to the
transport equation \citep{hbjcam99}.  In addition to the NLTE
transitions, all other LTE line opacities for atomic species not
treated in NLTE are treated with the equivalent two-level atom source
function, using a thermalization parameter, $\alpha =0.05$.  The
atmospheres are iterated to energy balance in the co-moving frame;
while we neglect the explicit effects of time dependence in the
radiation transport equation, we do implicitly include these effects,
via explicitly including the rate of gamma-ray deposition in the generalized
equation of radiative equilibrium and in the rate equations for the
NLTE populations.

The models are parameterized by the time since explosion and the
velocity where the continuum optical depth in extinction at
5000~\ang\ ($\tstd$) is unity, which along with the density profile
determines the radii. This follows since the explosion becomes
homologous ($v \propto r$) quickly after the shock wave traverses the
entire star. The density profile is taken to be a power-law in radius:
\[ \rho \propto r^{-n} \]
where $n$ typically is in the range $6-10$. Since we are only modeling
the outer atmosphere of the 
supernova, this simple parameterization agrees well with detailed
simulations of the light curve \citep{blinn87a00} for the relatively
small regions of the ejecta that our models probe.

Further fitting parameters are the model temperature $\Tmod$, which is
a convenient way of parameterizing the total luminosity in the
observer's frame. We treat the $\gamma$-ray deposition in a simple
parameterized way, which allows us to include the effects of nickel
mixing which is seen in nearly all SNe~II.
Detailed fitting of a time series of the observed spectra determines
all the parameters, i.e., $n$, $\Tmod$, the amount of nickel mixing,
the amount of helium mixing, and the metallicity.

\section{Results}
SN 1993W was discovered on Aug. 19, 1993 by C.~Pollas (IAUC
5848). Spectroscopy was obtained on Aug. 20 and Aug. 24 at ESO; and on
Sept. 10, Sept. 11, Oct. 22, and Nov. 8 at Lick Observatory.

\begin{figure}
\includegraphics[width=12cm,angle=90]{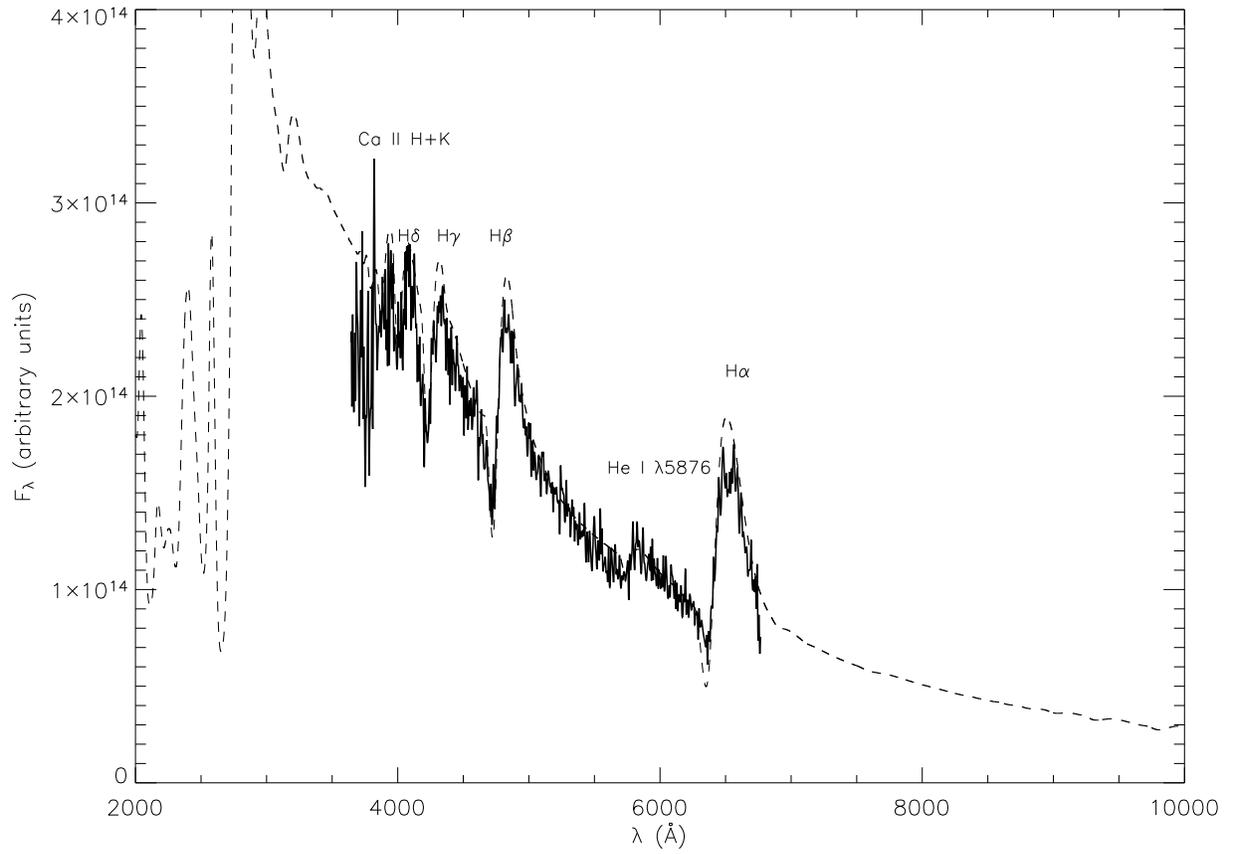}
\caption{\label{fig:aug20bestfit} The best fit model for the observed spectrum
obtained on Aug 20, 1993 using the ESO 3.6~m telescope and EFOSC.
The exposure time was 10 min.}
\end{figure}

\begin{figure}
\includegraphics[width=12cm,angle=90]{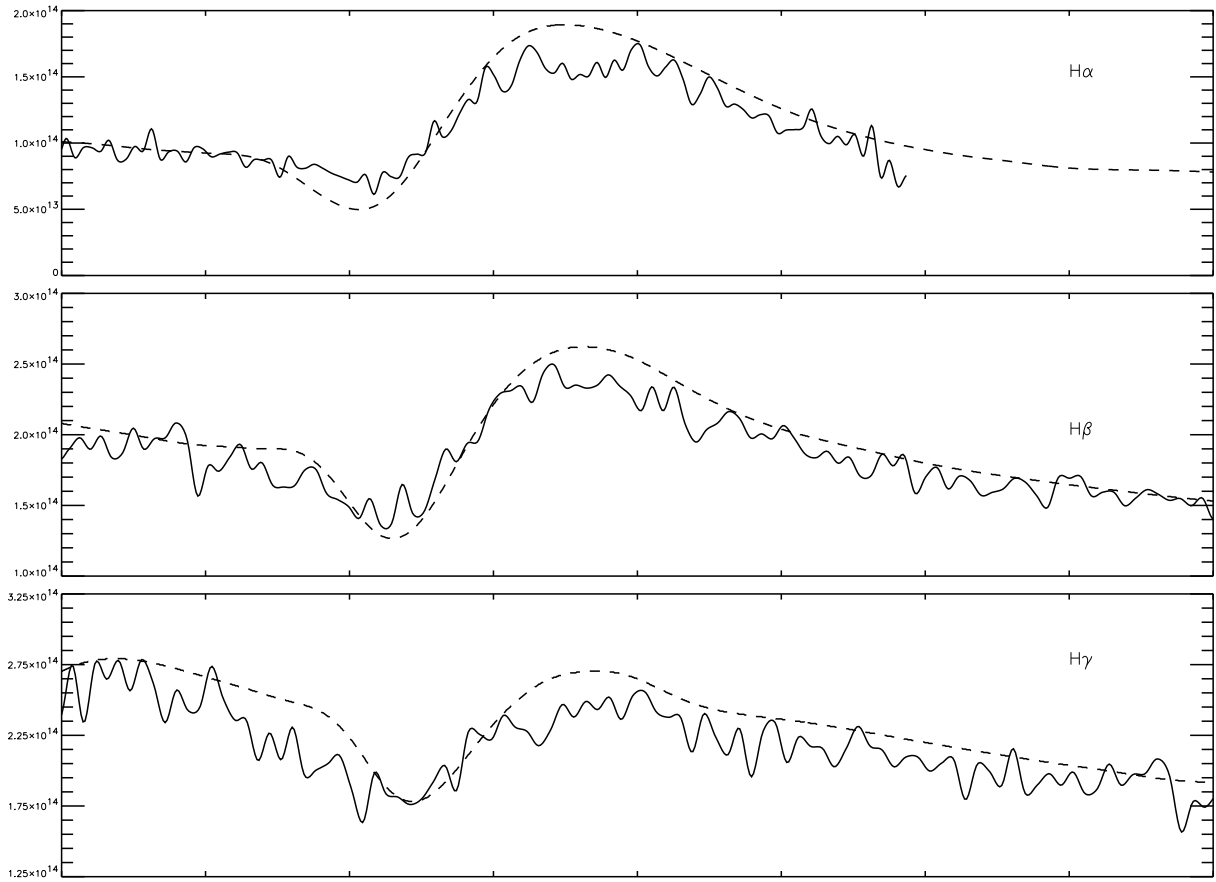}
\caption{\label{fig:aug20vel} The Balmer lines for Aug 20, plotted as
a function of velocity. The velocity structure is very well reproduced
by the synthetic spectra.}
\end{figure}

\begin{figure}
\includegraphics[width=12cm,angle=90]{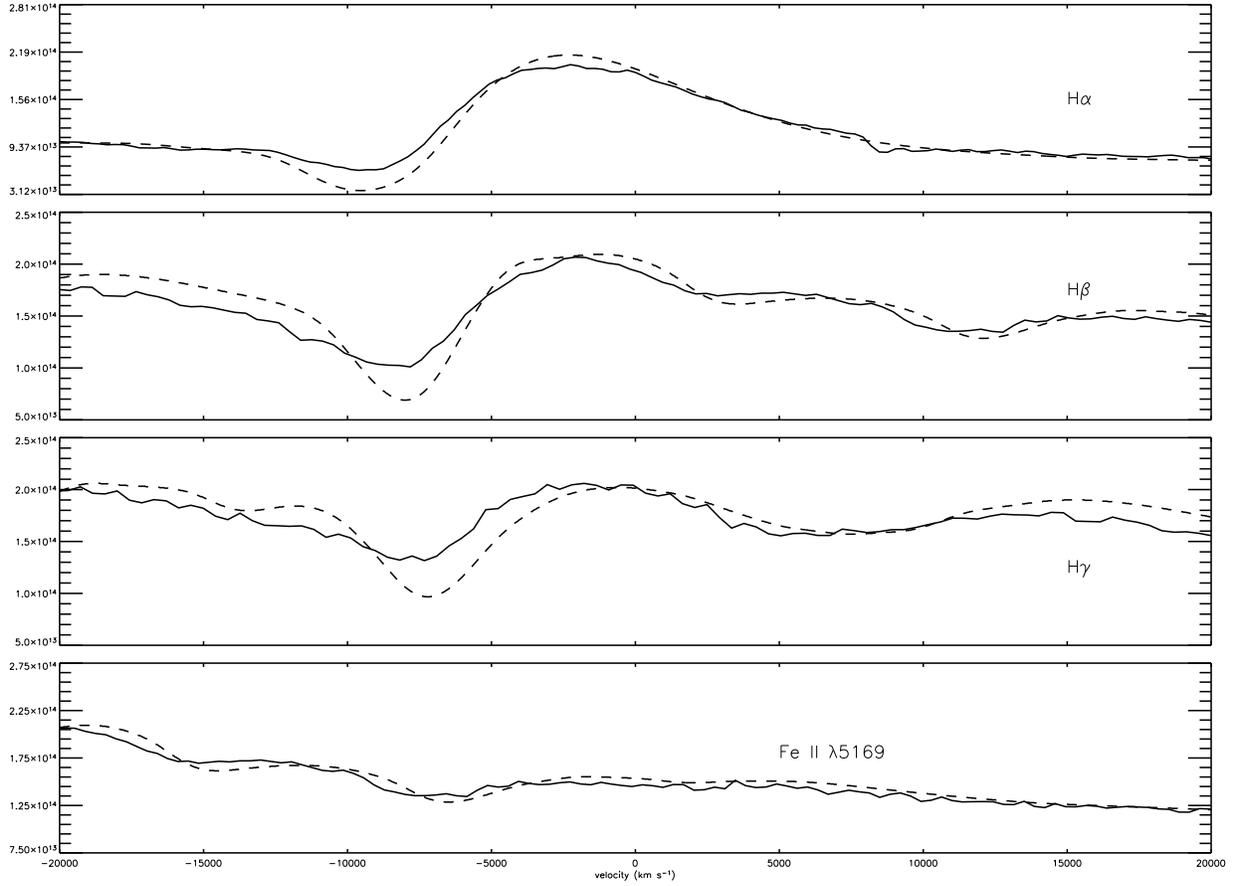}
\caption{\label{fig:aug24vel} The Balmer lines and the Fe~II
$\lambda$5169 line for Aug 24, plotted as
a function of velocity. The velocity structure is very well reproduced
by the synthetic spectra. In particular the weak Fe~II line is well
fit, which shows that the detailed modeling should be accurate for
SEAM based distance determinations (see text).}
\end{figure}

Figures~\ref{fig:aug20bestfit} and \ref{fig:aug24bestfit} display the observed
and best fit synthetic spectra for the two early epochs that we will
study here. The observed spectra were de-redshifted by 5400~\kmps\ and
de-reddened using the reddening law of \citet{card89} with
$E(B-V)=0.047$, i.e., accounting for galactic foreground
extinction. There is no evidence of interstellar reddening in the
parent galaxy since no narrow Na~ID lines are visible in the observed
spectra. Table~\ref{tab:params} displays the model parameters used for 
each date. The fit quality is superb, no line deviates from the
observed feature by more than 5\% in the Aug 20 spectrum and the
deviation is below 50\% (except in Ca II which we discuss below) in
the Aug 24 spectrum, which is 
excellent. Table~\ref{tab:syncolors} displays the synthetic colors for
the best fit models and Table~\ref{tab:obcolors} lists the observed
values. For Aug.~24, the synthetic $B-V$, agrees with the 
observed colors to within 0.04~mag which is within the observational
uncertainties. The $U-B$ colors differ somewhat more, but $U$
photometry is often uncertain and when the Ca H+K line is well fit the
synthetic value of $U-B = 0.20$ (see
Figure~\ref{fig:aug24low_ca}). Clearly, our ability to fit 
observed spectra in such detail gives us confidence in the physical
parameters that we can extract from our models. 
Figure~\ref{fig:aug20vel}  displays the Balmer
lines as a function of velocity for Aug 20, 1993 and
Figure~\ref{fig:aug24vel} displays both the Balmer lines and the weak
Fe~II $\lambda$5169 line which is often used in the expanding
photosphere method
\citep{schmkireas92,esk96,schmkireas92,hamuyepm01,leonard99em02}. This
result shows that the Spectral-fitting Expanding Atmosphere method
\citep[SEAM,][]{b93j1,b93j2,b93j3,b93j4,b94i1,mitchetal87a02} fits not
only the weak lines such as Fe~II $\lambda$5169 but also the prominent
SNe~II Balmer lines. The velocities determined by
detailed spectral analysis are extremely accurate (close to the
instrumental resolution) and indicates that
early spectra are quite important for SEAM analyses.

\begin{figure}
\includegraphics[width=12cm,angle=90]{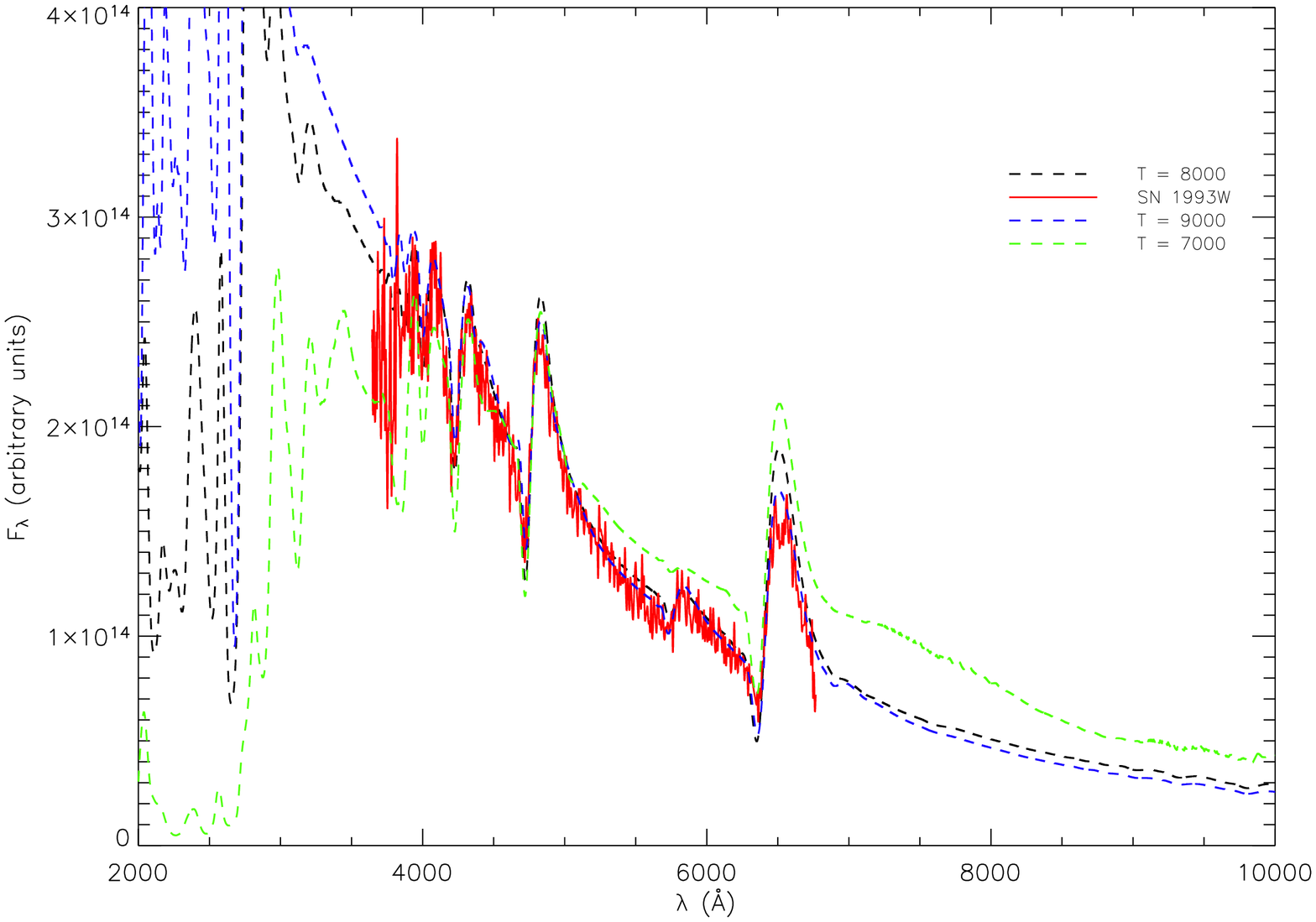}
\caption{\label{fig:aug20tvary} The  effect of varying the
model temperature $\Tmod$ is shown for the spectrum obtained on 
Aug 20, 1993.}
\end{figure}

\begin{figure}
\includegraphics[width=12cm,angle=90]{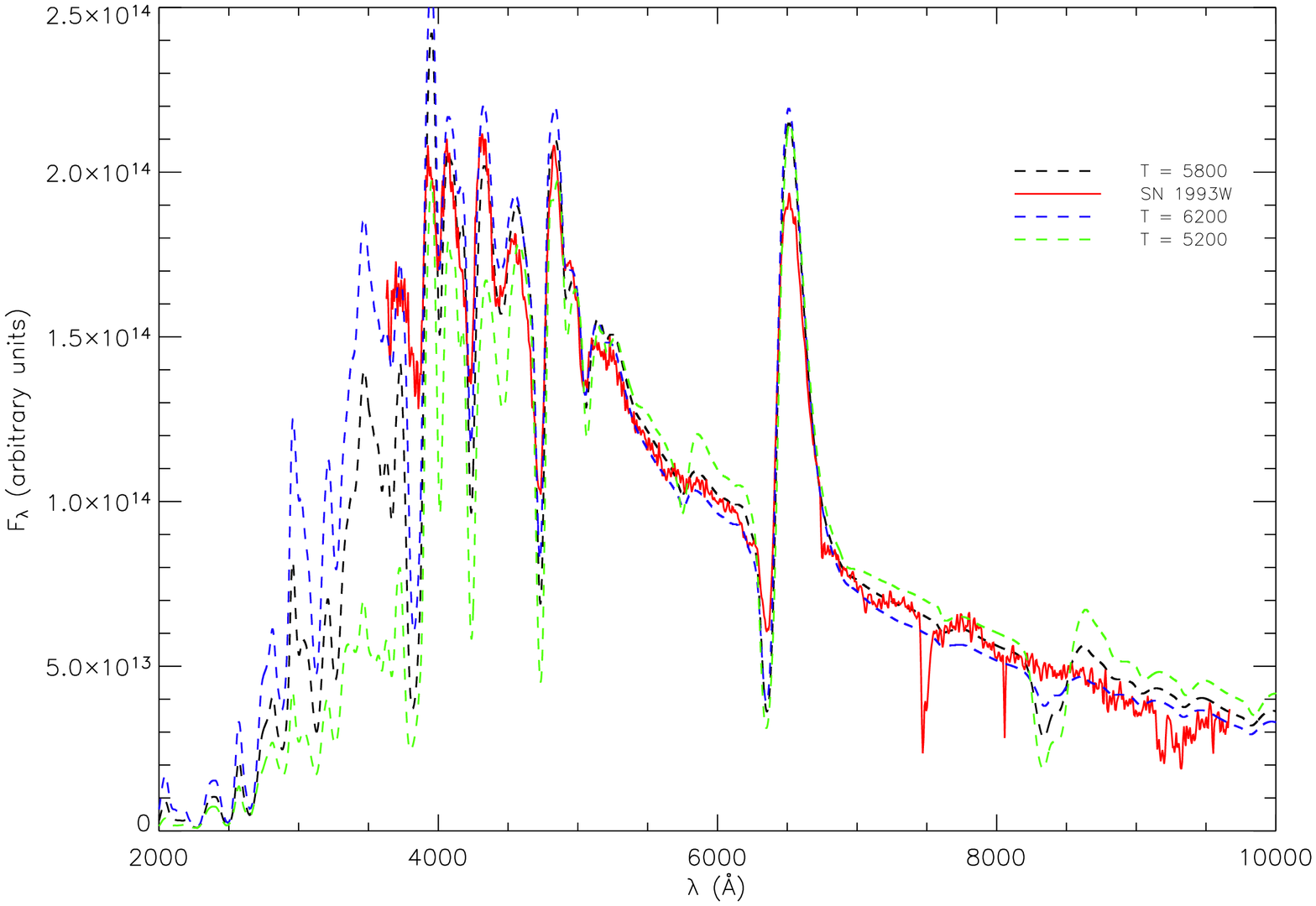}
\caption{\label{fig:aug24tvary} The  effect of varying the
model temperature $\Tmod$ is shown for the spectrum obtained on 
Aug 24, 1993.}
\end{figure}

Besides velocity, the other important fitting parameter is the total
observed bolometric luminosity which provides the outer boundary
condition for the requirment of global energy conservation. It is
convenient to parameterize the luminosity in terms of 
the model temperature \Tmod. Figures~\ref{fig:aug20tvary} and
\ref{fig:aug24tvary} show the effect of varying the the model
temperature on the standard reference models. From
Figure~\ref{fig:aug20tvary} it is clear that the 7000~K model is too
cool, producing both too little flux in the blue and far too much in
the red. The 9000~K model produces too much flux in the blue, but with
the spectral coverage of the observation, it is difficult to
distinguish it in the red from our standard 8000~K
model (see Fig.~\ref{fig:aug20bestfit}). Thus we can obtain the model
temperature at this single epoch 
to $\pm 500$~K. With larger 
spectral coverage we should be able to determine the model temperature
to $\sim \pm 100$~K. Fig.~\ref{fig:aug24tvary} shows that on Aug 24, the
cool model produces poorer fits to the 
Balmer lines, produces a very strong Ca~II IR triplet (which also
depends somewhat on the metallicity, discussed below) and also does
less well at reproducing the Ca~II H+K line. The hotter model also does a
poorer job of reproducing the Balmer lines, and although the Ca~II IR
triplet is weaker, as needed by the observations, the Ca~II H+K line
is still far too strong. 
 
\subsection{Aug 20}

\begin{figure}
\includegraphics[width=12cm,angle=90]{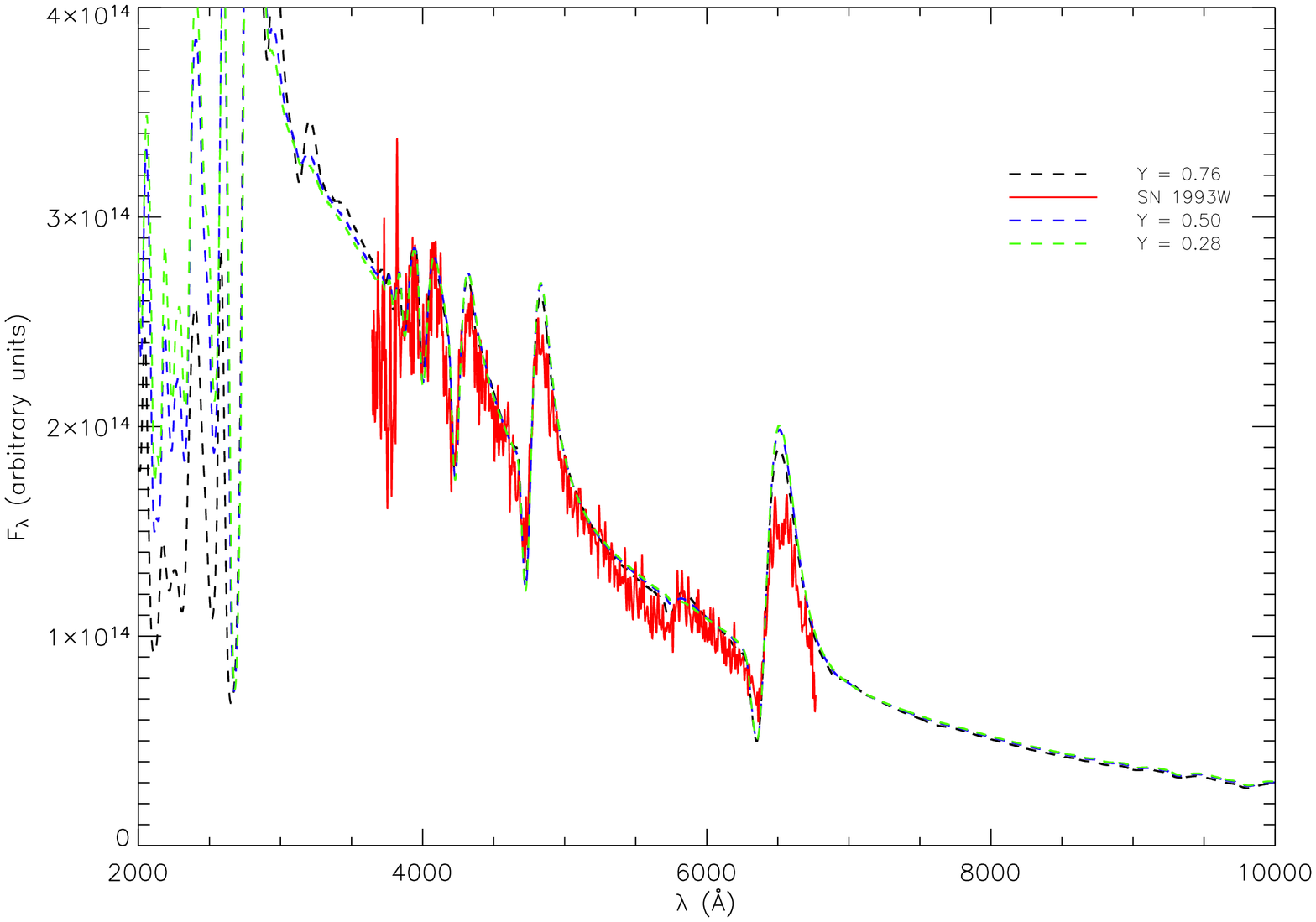}
\caption{\label{fig:aug20hevary} The  effect of varying the
helium abundance is shown for the spectrum obtained on 
Aug 20, 1993.}
\end{figure}

\begin{figure}
\includegraphics[width=12cm,angle=90]{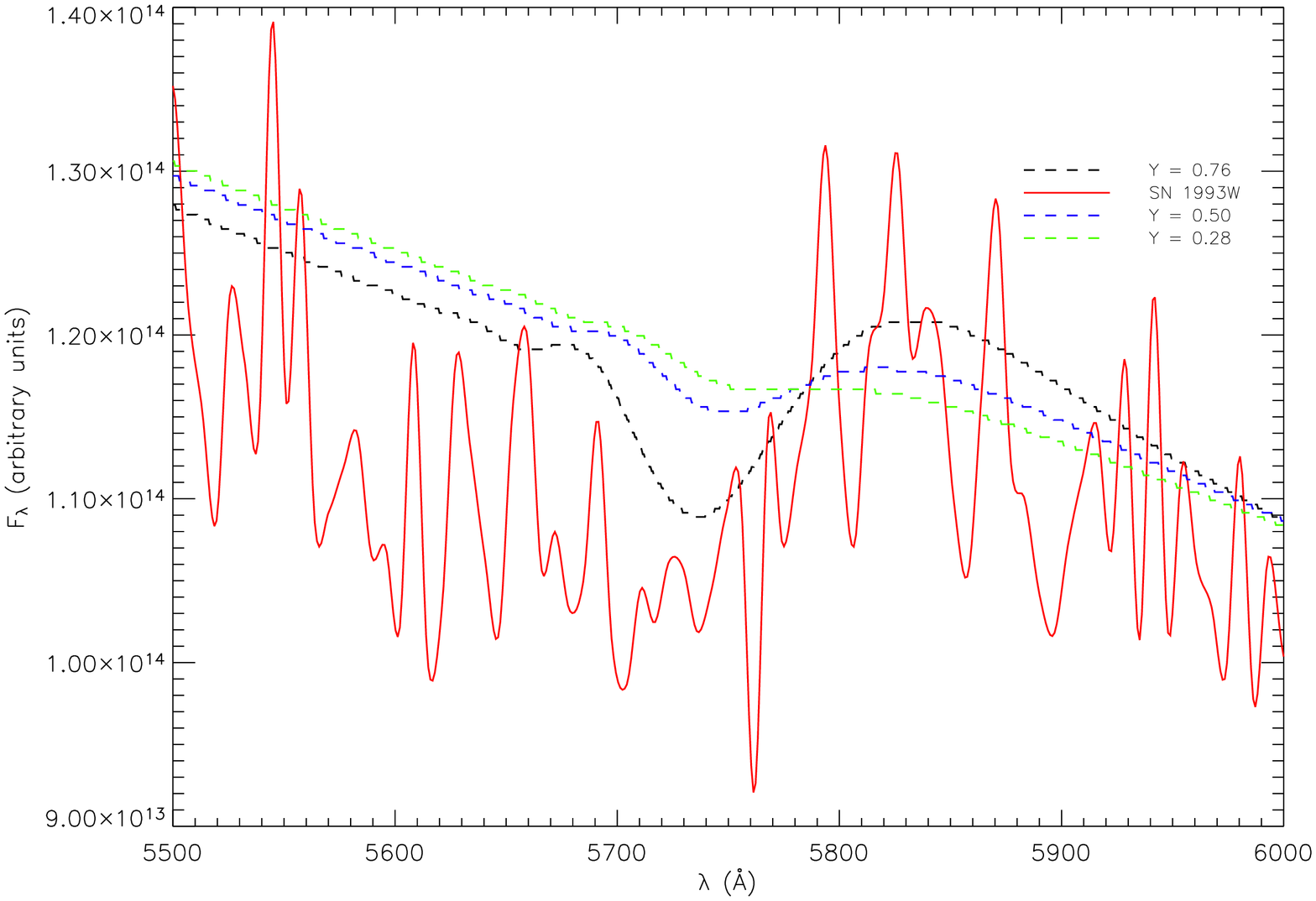}
\caption{\label{fig:aug20hevary_2} The  effect of varying the
helium abundance on the He I $\lambda 5876$ line is shown for the
spectrum obtained on  
Aug 20, 1993.}
\end{figure}

\begin{figure}
\includegraphics[width=12cm,angle=90]{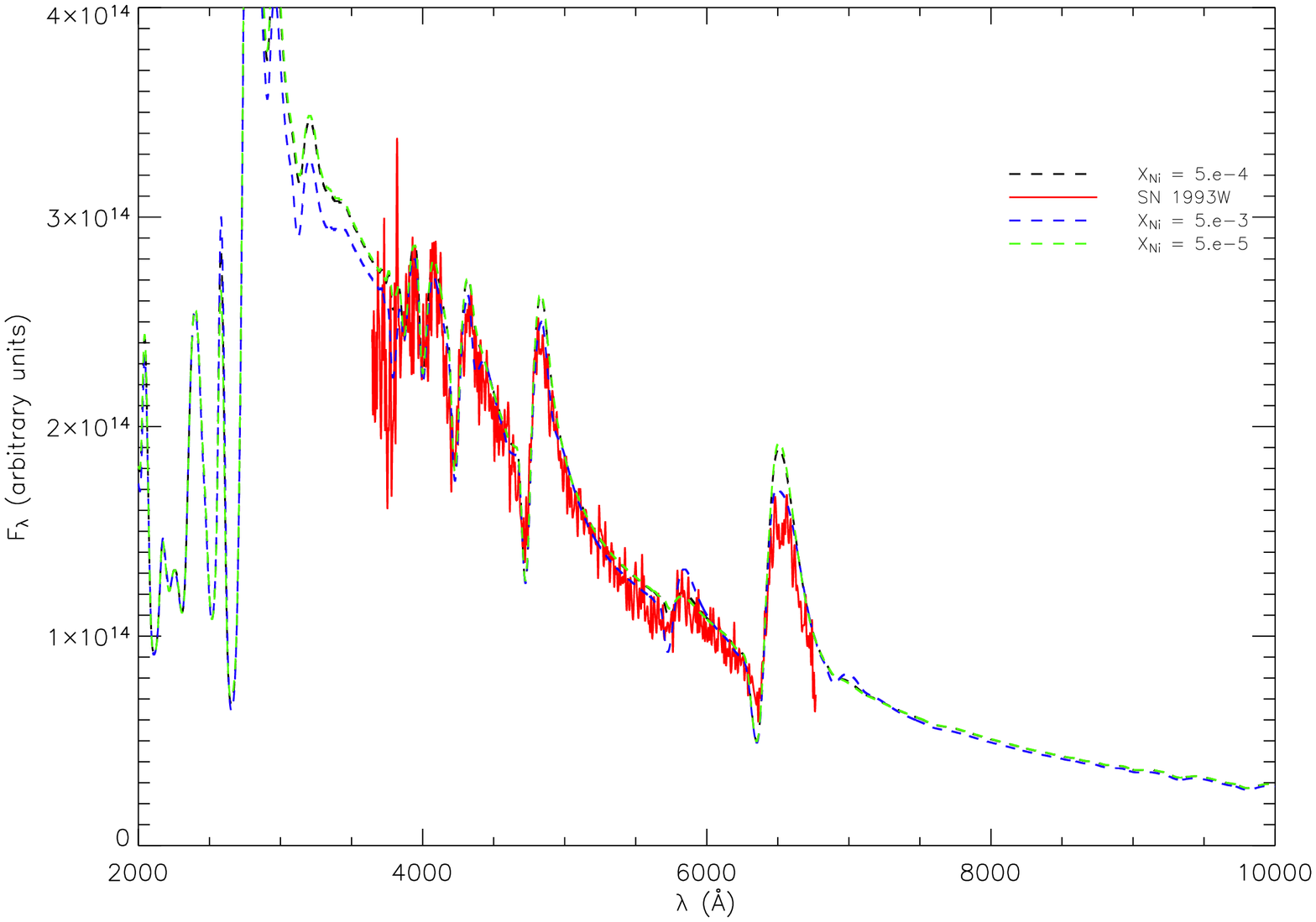}
\caption{\label{fig:aug20nivary} The effect of varying the
nickel mixing is shown for the spectrum obtained on 
Aug 20, 1993.}
\end{figure}

\begin{figure}
\includegraphics[width=12cm,angle=90]{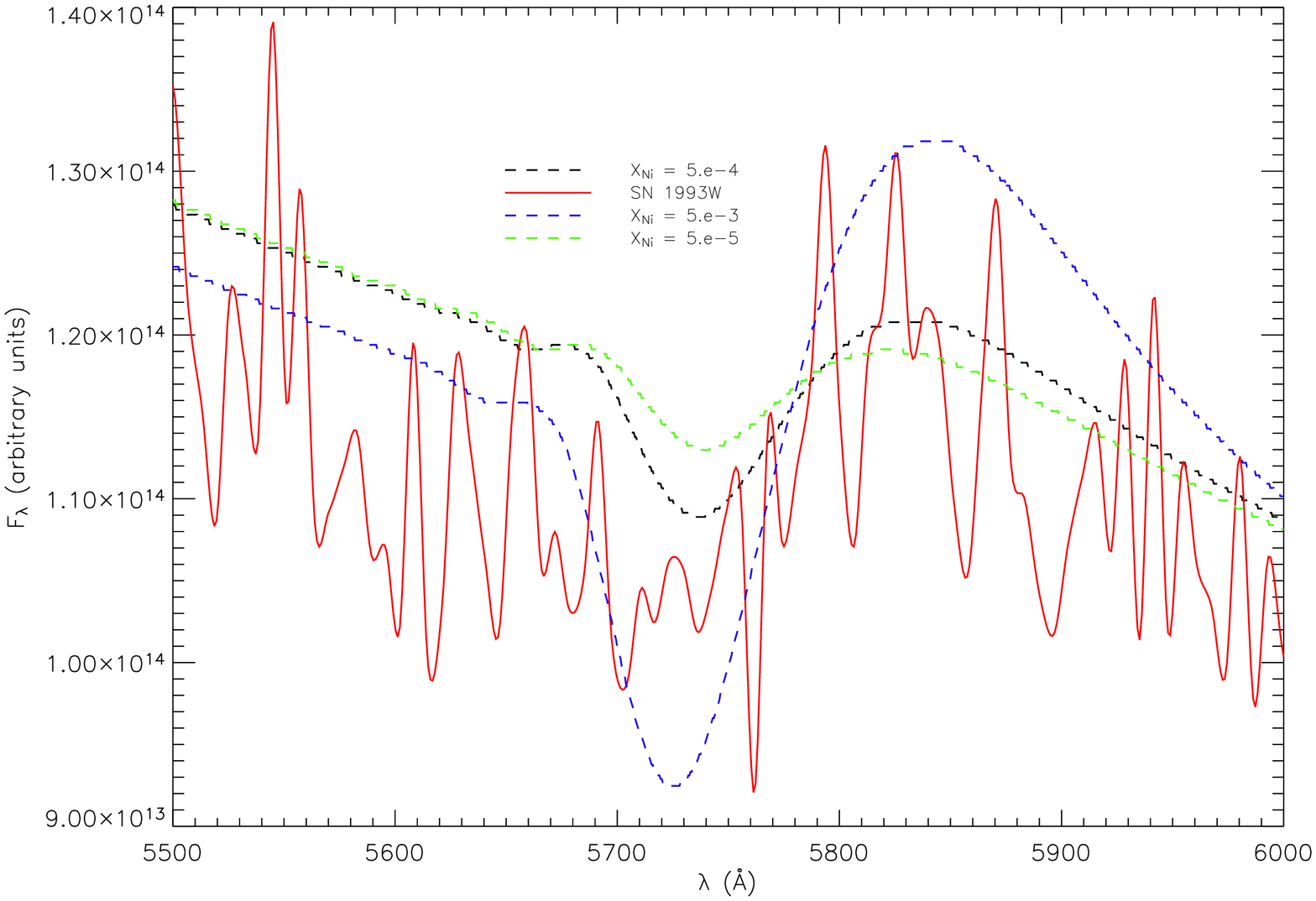}
\caption{\label{fig:aug20nivary_he} The effect of varying the
nickel mixing on the He I $\lambda 5876$ line is shown for the spectrum
obtained on  
Aug 20, 1993.}
\end{figure}

Because the Aug 20 spectrum is so hot and so early, it contains (in
the optical) only lines of hydrogen and helium. Thus we may use this
spectrum to determine the amount of hydrogen and helium mixing,
although we must simultaneously determine the amount of non-thermal
excitation due to the radioactive decay of \nni. We have calculated a
very large number  of models ($\sim 500$) and we find that no
reasonable value of 
$\gamma$-ray deposition can produce the observed He~I $\lambda5876$
line without enhancing the helium abundance over the solar value.  We
find that in order to fit this feature there must significant mixing
of hydrogen and helium in the outer envelope and the helium must also
be excited by non-thermal $\gamma$-ray deposition which comes from the
radioactive decay of \nni.  A similar result was found for SN 1987A
\citep{sn87arev89,eastkir89,LF87A96,blinn87a00,mitchetal87a01,mitchetal87a02}
and for SN~1999em \citep{bsn99em00}.  Figures
Figure~\ref{fig:aug20hevary} and \ref{fig:aug20hevary_2} show the
effects of varying the helium mass fraction. Figure~\ref{fig:aug20hevary_2}
shows that with a fixed amount of $\gamma$-ray deposition, solar abundances of
helium do not produce a strong enough He~I $\lambda5876$
feature. Figures~\ref{fig:aug20nivary} and \ref{fig:aug20nivary_he} show
the effect of varying the $\gamma$-ray deposition, with a constant
$Y=0.76$.  Even with enhanced 
helium abundances in the outer layers, the He~I $\lambda
5876$ feature does not appear with no non-thermal ionization,
therefore mixing of nickel (in order to 
produce $\gamma$-rays to excite the helium) into the outer layers is
required. The mixing of nickel, hydrogen, and helium in SNe~II is not
well understood and the situation is unlikely to improve until the
explosion mechanism is 
better understood. Therefore we use a very simple prescription for
$\gamma$-ray deposition, the $\gamma$-rays are deposited locally by a
constant mass fraction of nickel \xni\ --- the deposition function
follows the density profile. We have used this prescription
successfully in modeling other SNe~II
\citep{l98s01,mitchetal87a01,mitchetal87a02}. Both enhanced surface
helium 
and the excitation of the helium by $\gamma$-ray deposition appear to
be a common feature of SNe~IIP and support the notion that fluid
instabilities or MHD jets are intimately involved in the explosion
mechanism in order to produce large amounts of mixing.

\begin{figure}
\includegraphics[width=12cm,angle=90]{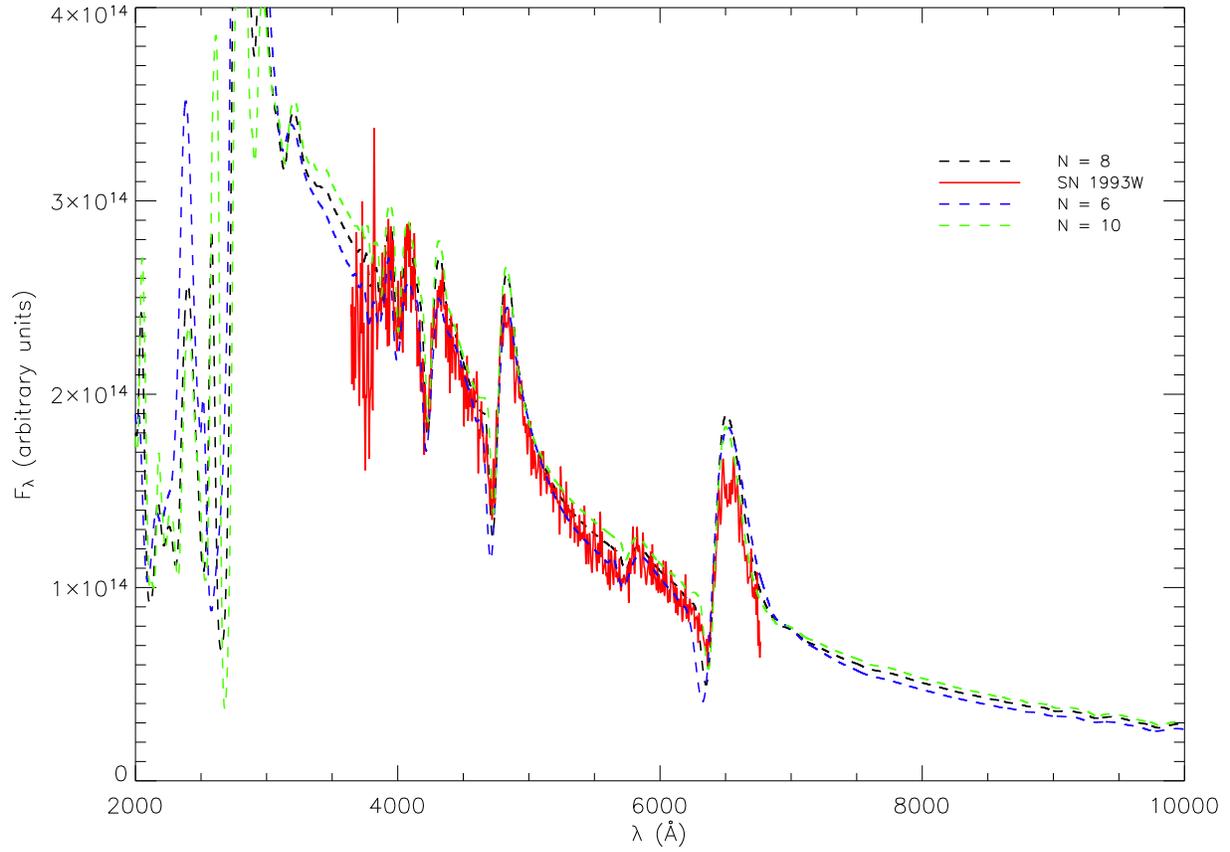}
\caption{\label{fig:aug20nvary} The  effect of varying the
density profile $n$ is shown for the spectrum obtained on 
Aug 20, 1993.}
\end{figure}

Finally, Figure~\ref{fig:aug20nvary} displays the effects of varying
the density profile $n$. The Balmer profiles are too narrow in the
$n=10$ case and the $n=6$ model does nearly as well as our standard
$n=8$ case. There is some degeneracy between the value of $n$ and the
velocity, since the narrow lines in the $n=10$ case could be made
broader by increasing the velocity of the ejecta, however in our large
grid of models we found that lower values of $n$ are favored, largely
due to the fact that changing $n$ and increasing the velocity alters
the slope of the P-Cygni feature as it rises from the bottom of the
absorption trough to the emission peak.

\subsection{Aug 24}

\begin{figure}
\includegraphics[width=12cm,angle=90]{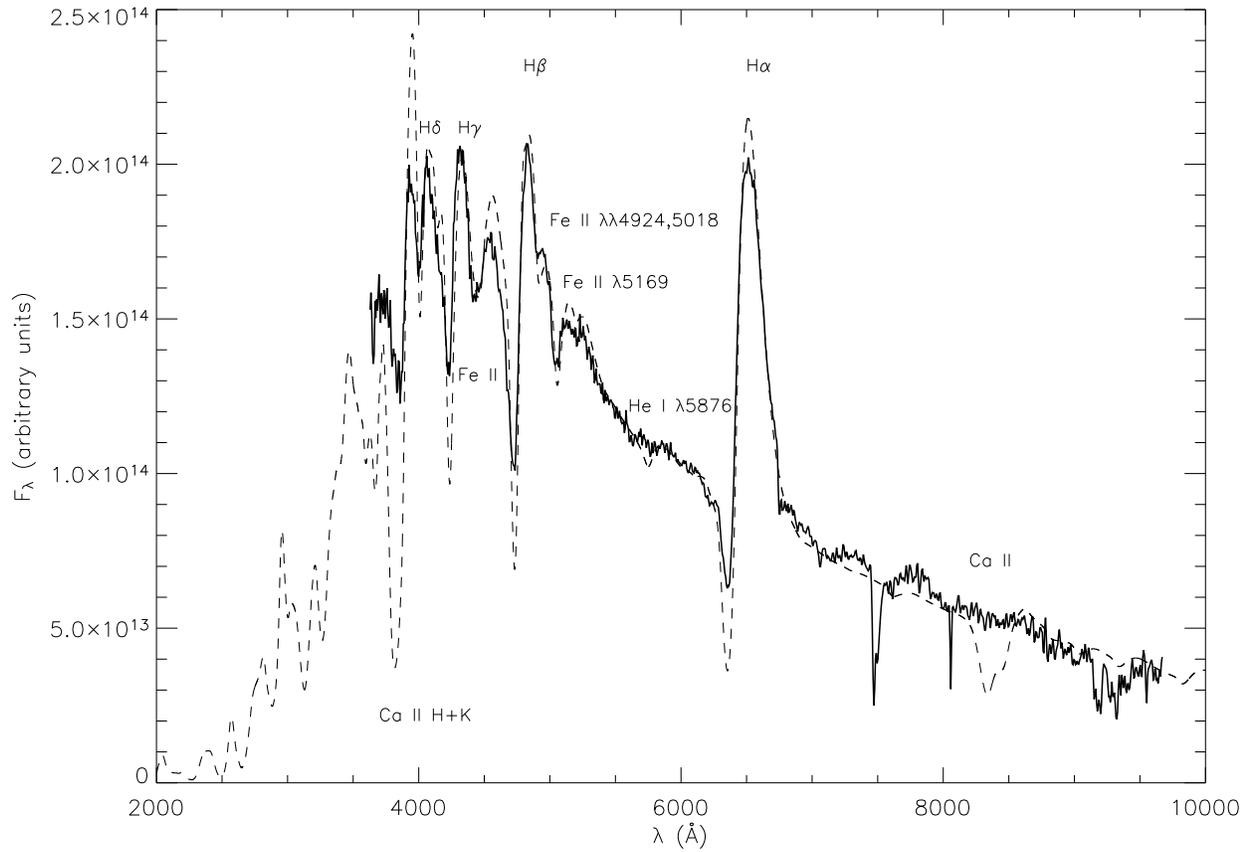}
\caption{\label{fig:aug24bestfit} The best fit model for the observed spectrum
obtained on Aug 24, 1993 also using ESO 3.6~m and EFOSC. To extend the
wavelength coverage we merged two spectra obtained with different grisms.
The total exposure time was 30 min.}
\end{figure}

\begin{figure}
\includegraphics[width=12cm,angle=90]{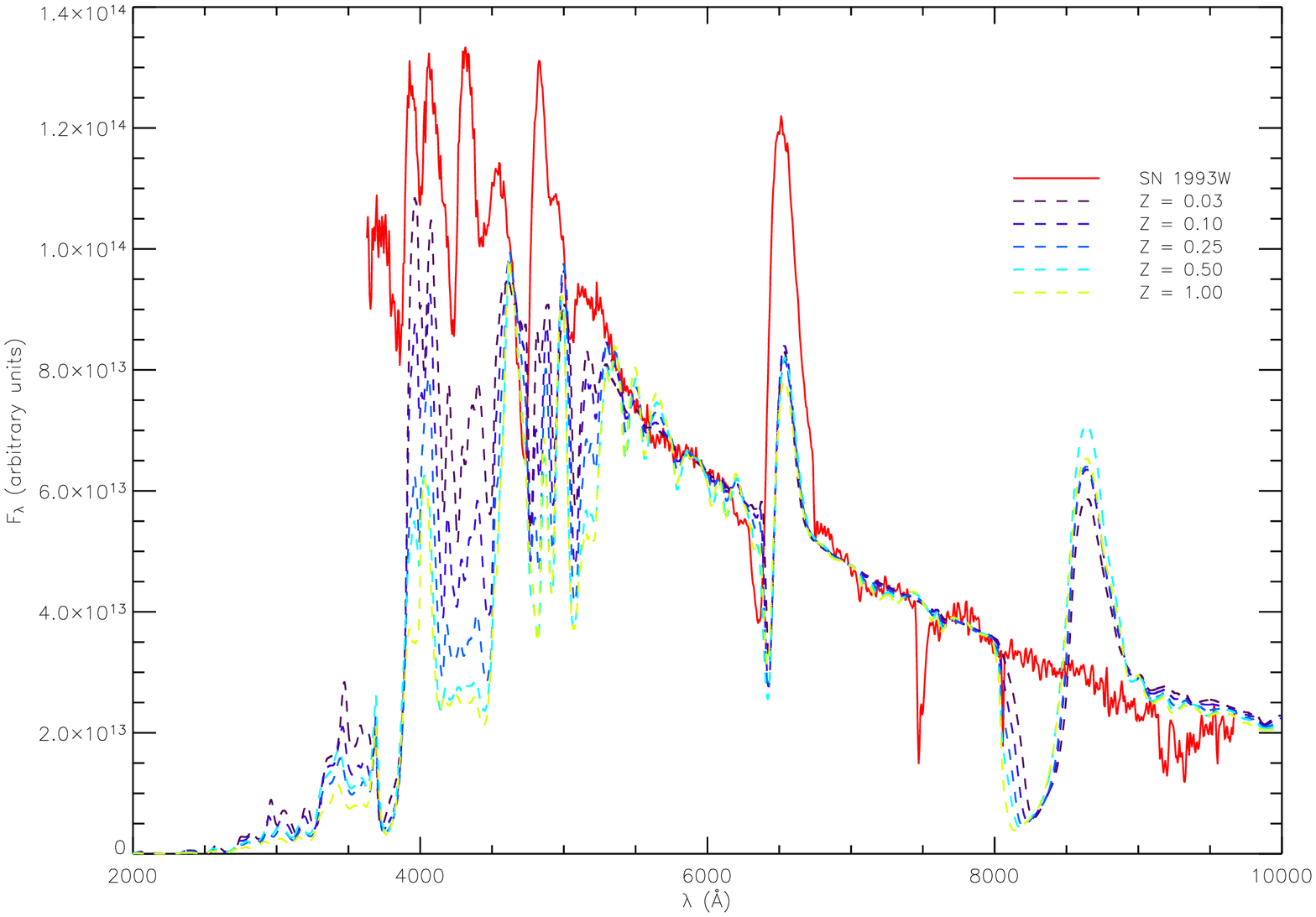}
\caption{\label{fig:aug24znoni} The  effect of varying the
metal abundance is shown for the spectrum obtained on 
Aug 24, 1993. No $\gamma$-ray deposition is included in these
calculations.}
\end{figure}

\begin{figure}
\includegraphics[width=12cm,angle=90]{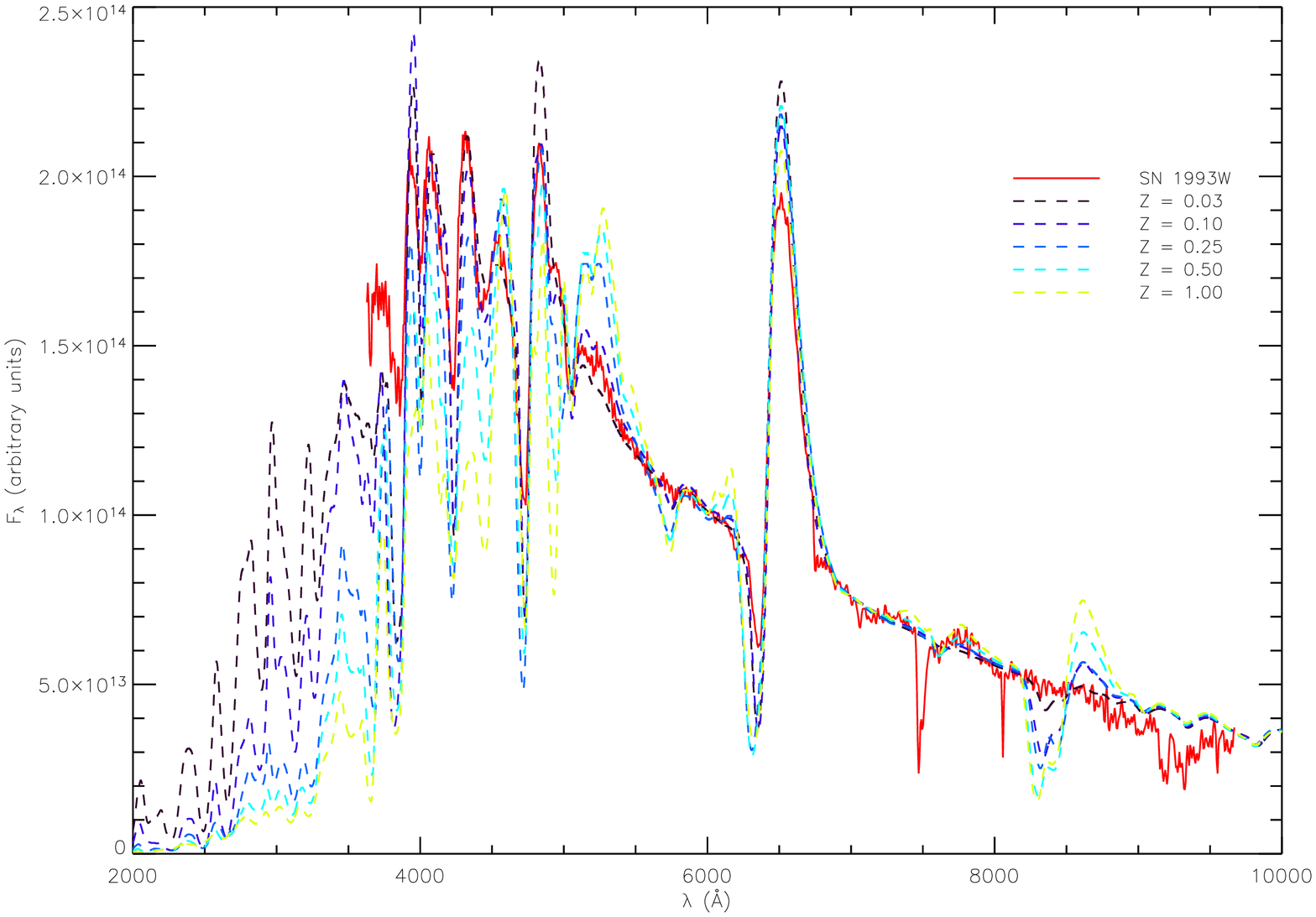}
\caption{\label{fig:aug24zwni} The  effect of varying the
metal abundance is shown for the spectrum obtained on 
Aug 24, 1993. $\gamma$-ray deposition is included in these
calculations.}
\end{figure}

\begin{figure}
\includegraphics[width=12cm,angle=90]{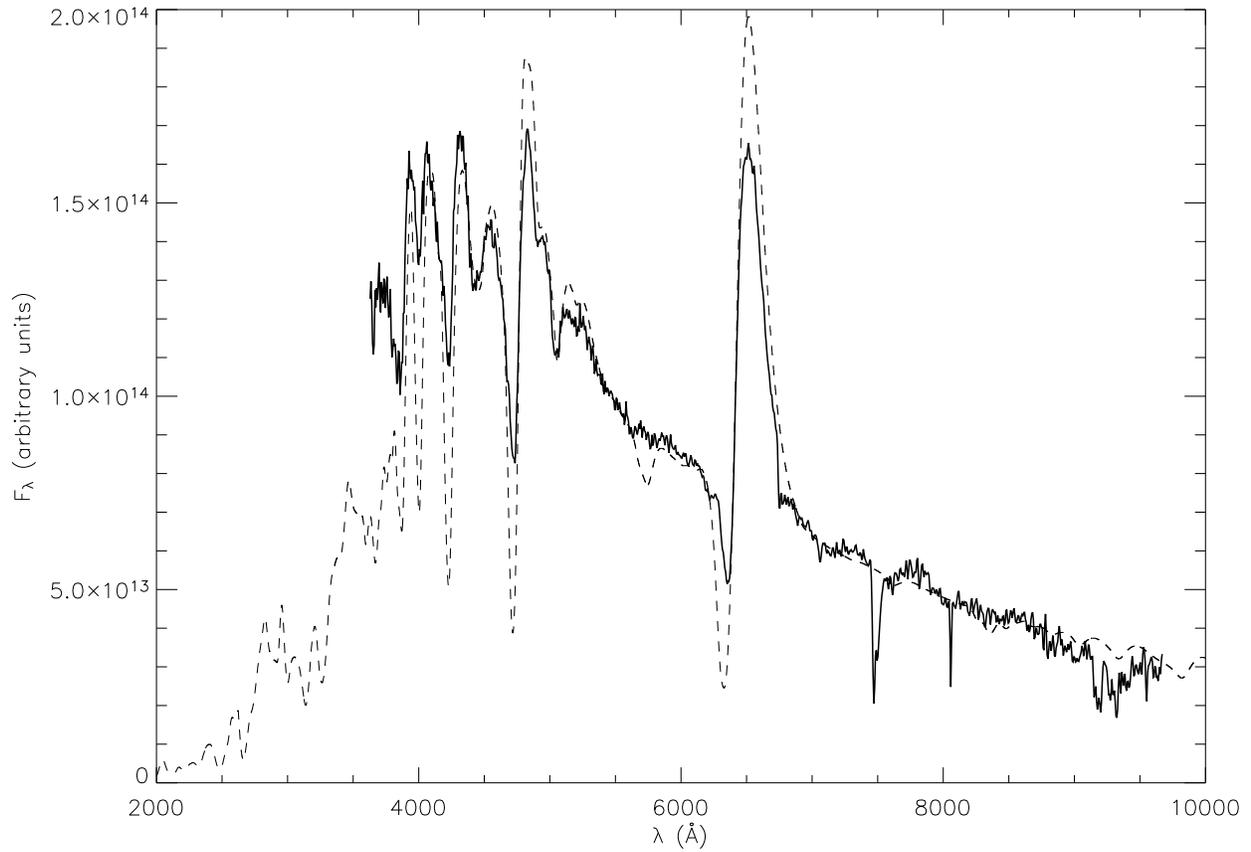}
\caption{\label{fig:aug24low_ca} The synthetic spectrum with the same
parameters as that of the best fit model (Fig.~\ref{fig:aug24bestfit})
except that the calcium abundance has be reduced by a factor of ten is
compared with the observed spectrum
obtained on Aug 24, 1993.}
\end{figure}

Figure~\ref{fig:aug24bestfit} shows that the \ion{Fe}{2}
$\lambda5169$ line has now appeared in the observed spectrum and a
weak feature is also visible that is likely due to \ion{Fe}{2}
$\lambda\lambda4924,5018$. However, there is not significant line
blanketing in 
the blue even though the model temperature $\Tmod$ has dropped to the
hydrogen recombination temperature (see
Fig.~\ref{fig:aug24znoni}). Figure~\ref{fig:aug24znoni} shows 
the effect of varying the model metallicity with gamma-ray deposition
turned off. The results are expected, higher metallicity leads to large
absorption due to iron line blanketing in the region
4000--4500~\ang. In fact even at very low metallicity the line
blanketing is much larger than observed, gamma-ray deposition
significantly reduces the line blanketing as shown in
Figure~\ref{fig:aug24zwni}. In fact, the comparison of the two
figures strengthen the need for significant gamma-ray deposition at
early times. While Fig.~\ref{fig:aug24zwni} clearly shows that there
is more line blanketing as the metallicity increases, it also shows
that metallicity and $\gamma$-ray deposition radically alter the flux
in the blue part of the spectrum.

\begin{figure}
\includegraphics[width=12cm,angle=90]{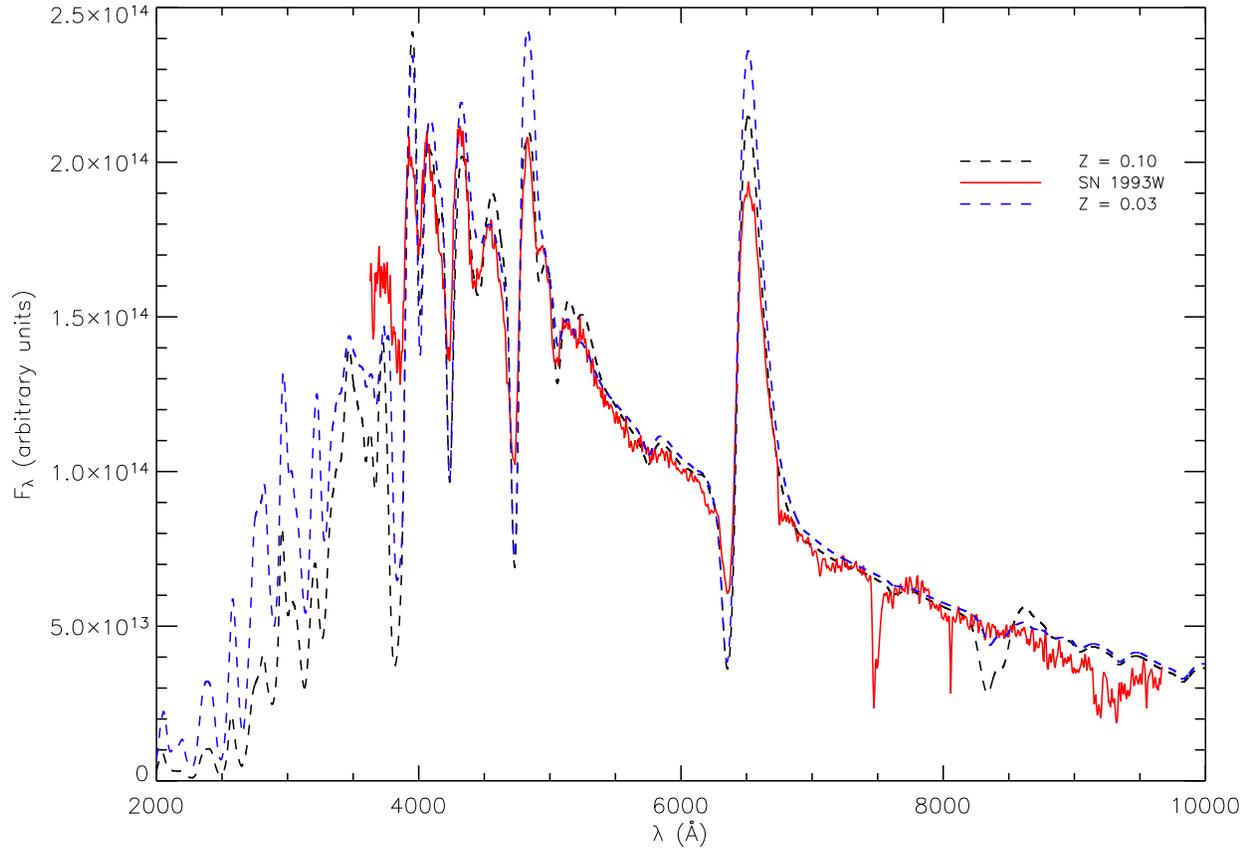}
\caption{\label{fig:aug24zvary} The  effect of varying the
metal abundance at low metallicity is shown for the spectrum obtained on 
Aug 24, 1993, $\gamma$-ray deposition is included.}
\end{figure}

\begin{figure}
\includegraphics[width=12cm,angle=90]{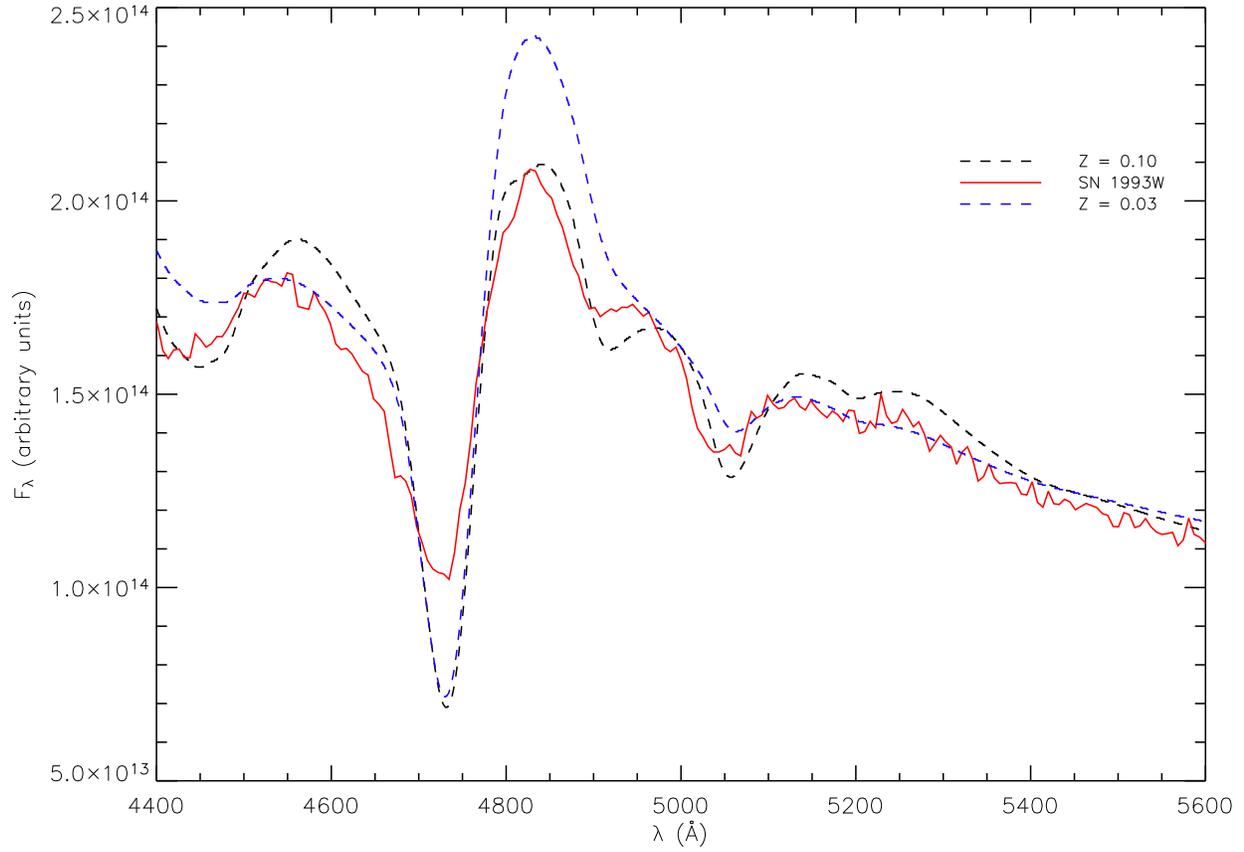}
\caption{\label{fig:aug24zvary_feii} The  effect of on the Fe II $\lambda
5169$ line and the Fe II $\lambda\lambda4924,5018$ lines of varying the
metal abundance at low metallicity is shown for the spectrum obtained on 
Aug 24, 1993, $\gamma$-ray deposition is included.}
\end{figure}

Figure~\ref{fig:aug24zvary} shows that even at very low metallicity,
the early spectra are also 
quite sensitive to 
the primordial metallicity,  when
gamma-ray deposition is 
included. The higher metallicity $Z=0.10Z_\odot$ is favored to fit the
two \ion{Fe}{2} features in the blue (see
Fig.~\ref{fig:aug24zvary_feii}), but the \ion{Ca}{2} IR triplet 
is too strong with $Z=0.10Z_\odot$, and is even a bit too strong for
$Z=0.03Z_\odot$. The \ion{Ca}{2} H+K feature is too strong in both
cases and is likely affected by the choice of gamma-ray deposition as
can be seen from  Figure~\ref{fig:aug24znoni}. Since we have not tried
to alter individual abundances, but 
only scaled total overall metallicity our preferred value for
metallicity is set by the iron features in the blue. Clearly the
progenitor star for SN~1993W 
had a significantly lower metallicity than the sun. 

Fig.~\ref{fig:aug24low_ca} shows a model with the same parameters as
Figure~\ref{fig:aug24bestfit}, but with the Ca abundance reduced by a
factor of 10. Now both the H+K and IR triplet lines are reasonably
well reproduced in the synthetic spectrum; however since we have held
everything else fixed the Balmer lines are not as well fit as they are
in Figure~\ref{fig:aug24bestfit}. It is interesting to note that
nucleosynthetic models for Type Ia and Type II supernovae from low mass
progentitors produce Fe/Ca
ratios that are about a factor of four higher than that produced by
more massive SNe~II
\citep{tsujetal95}. It is possible that the progenitor of SN~1993W
just happened to occur in a region that was recently enriched by the
explosion of a Type Ia supernova, resulting in enhanced
Fe/Ca. Although low S/Fe and Si/Fe would also be expected in this
scenario these are not ruled out by the spectra presented here. Si and
S abundances can best be determined in later (nebular) spectra of SNe
IIP. Since calcium lines are so strongly dependent on the temperature,
small changes in the non-thermal ionization could possibly produce the
suppression in the calcium strengths that we obtain here. A detailed
study of a full time series of a similar SN IIP would thus be
illustrative and we intend to carry out such a program in future work.

We find that
$0.03 < Z_{\mbox{SN 1993W}}/Z_\odot < 0.15$, and we strongly prefer a value of
$Z_{\mbox{SN 
1993W}}/Z_\odot \approx 0.10$ (particularly if Fe/Ca is enhanced). The
metallicity likely could  
be even better pinned down with UV observations that can be obtained
by \emph{HST} 
and by \emph{SNAP}. By studying the progenitors of SNe~II,
\citet{smartt99em02}, found that metallicities near solar with on
SN~1987A and SN~1980K having metallicities of about $0.5Z_\odot$. Thus
the value of $0.10Z_\odot$ for SN~1993W is interestingly low. Therefore
early observations of SNe~II are crucial if we 
wish to use these objects as probes of galactic chemical evolution.

\section{Explosion Time}

We can use simple arguments to obtain a limit on the explosion time.
The total bolometric luminosity of the supernova is approximately
\begin{eqnarray*}
L &=& 4\pi R^2 \sigma T^4\\
  &=& 4\pi v^2_\mathrm{ph} t^2 T^4
\end{eqnarray*}
where $v_\mathrm{ph}$ is the velocity of the photosphere. The ratio of
the luminosities between time $t_1$ (Aug 20) and time $t_2$ is

\begin{eqnarray*}
\frac{L_1}{L_2} &=&  \frac{v^2_\mathrm{ph_1} t_1^2 T_1^4}{v^2_\mathrm{ph_2}
t_2^2 T_2^4}\\
 &=&  \frac{v^2_\mathrm{ph_1} t_1^2 T_1^4}{v^2_\mathrm{ph_2}
(t_1+3.90)^2 T_2^4}
\end{eqnarray*}
where we have used the fact that $t_2=t_1+3.90$~days. From our synthetic
spectra we can determine that the bolometric correction was $BC =
-0.40$ on Aug 20, and $BC = 0.04$ on Aug 24. The $B$ magnitude is
estimated to be 18.3 on both days (see Tab.~\ref{tab:obphoto}) and
using the $B-V$ color from 
Table~\ref{tab:syncolors} we find that the ratio of $L_1$ to $L_2$ is
1.17. Plugging in our values for the velocities and temperatures we
find that  Aug 20 is  $4.7 \pm 0.7$~days after the explosion. The
largest contributor to the uncertainty in the explosion time is the
error in the 
observed photometry.

\section{Conclusions}
We have shown that synthetic spectra of SNe~IIP can be fit to high
accuracy, and that the primordial metallicity, degree of mixing of H
and He, as well as the amount of mixing of \nni\ can also be
determined by early spectra. \citet{bsn99em00} showed that
the total amount of extinction due to dust either in the parent galaxy
or in our own galaxy can also be estimated with UV+optical spectra at
very early times. While the reduced calcium abundance that was needed
to reduce the strength of the calcium lines is intriguing, we do not
as yet believe that our results require a reduced value of Ca/Fe.  In
a future paper we will explore a grid of models 
and determine a distance to SN~1993W, along with an estimate of the
systematic error. We have shown that primordial metallicity, H/He
mixing, and \nni mixing are derivable from the detailed analysis of
early spectra of SNe IIP. The analysis of abundances of individual
elements is in reach, but will require a larger grid of models
intercompared with a set of well studied SNe IIP. The data for this
will be obtained as part of searches for nearby SNe~Ia and will just
require followup spectrophotometry. We believe that we
have demonstrated that SNe~IIP atmospheres can be well understood
theoretically and hence are likely to make excellent independent cosmological
probes, both as distance indicators and as probes of cosmic
nucleosynthesis. 

\acknowledgments
PHH was supported in part by the P\^ole Scientifique de Mod\'elisation
Num\'erique at ENS-Lyon. This work was supported in part by NASA
grant NAG5-12127, NSF grant AST-0204771, and an IBM SUR 
grant to the University of Oklahoma; and by NSF grant AST-9720704,
NASA ATP grant NAG 5-8425 and LTSA grant NAG 5-3619 to the University
of Georgia. MT aknowledges the support of the Italian
Ministry of University
and Research through grant COFIN MM02905817.
 Some of the calculations presented in this paper were
performed at the San Diego Supercomputer Center (SDSC), supported by
the NSF, and at the National Energy Research Supercomputer Center
(NERSC), supported by the U.S. DOE. We thank both these institutions
for a generous allocation of computer time.  This research has made
use of the NASA/IPAC Extragalactic Database (NED) which is operated by
the Jet Propulsion Laboratory, California Institute of Technology,
under contract with the National Aeronautics and Space Administration.


\clearpage

\begin{deluxetable}{rrrrr}
\tablecolumns{5}
\tablewidth{0pc}
\tablecaption{Observed Photometry\label{tab:obphoto}}
\tablehead{
\colhead{Date}&  \colhead{$U$} & \colhead{$B$}& \colhead{$V$} & \colhead{$R$}}
\startdata
Aug 24&18.58&  18.32& 18.05&\nodata\\
Sep 09&\nodata&19.15& 18.25&17.95
\enddata
\end{deluxetable}

\begin{deluxetable}{rrrrrr}
\tablecolumns{6}
\tablewidth{0pc}
\tablecaption{Model Parameters\label{tab:params}}
\tablehead{
\colhead{Date}& \colhead{$\vno$}& \colhead{$n$} & \colhead{$\Teff$}&
\colhead{$Z$/$Z_\odot$} &\colhead{$Y$}}
\startdata
Aug. 20& 9356& 8& 9356& 0.10& 0.76\\
Aug. 24& 6897& 8& 6689& 0.10& 0.76
\enddata
\tablecomments{Model parameters for our best fit models shown in
Figures~\protect\ref{fig:aug20bestfit}--\ref{fig:aug24bestfit}. Date,
refers to the date of observation, $\vno$ is the velocity (in \kmps)
at the reference optical depth $\tstd=1$ ($\tstd =$ total extinction
optical depth in the continuum at 5000~\ang), $n$ is the exponent of
the density power-law, $\Teff$ characterizes the total luminosity of
the model, $Z/Z_\odot$ is the metallicity in units of the solar metallicity,
$Y$ is the fraction by mass that is helium.}
\end{deluxetable}

\begin{deluxetable}{rrrrrr}
\tablecolumns{5}
\tablewidth{0pc}
\tablecaption{Synthetic Colors\label{tab:syncolors}}
\tablehead{
\colhead{Date}& \colhead{$U-B$} & \colhead{$B-V$}& \colhead{$V-R$}&
\colhead{$R-I$}}
\startdata
Aug20&   -0.70 &   0.03 &   0.22 &   0.00\\
Aug24&    0.03 &   0.30 &   0.27 &   0.08
\enddata
\end{deluxetable}

\begin{deluxetable}{rrr}
\tablecolumns{3}
\tablewidth{0pc}
\tablecaption{Observed Colors\label{tab:obcolors}}
\tablehead{
\colhead{Date}& \colhead{$U-B$} & \colhead{$B-V$}}
\startdata
Aug24&0.26  &  0.26
\enddata
\end{deluxetable}

\clearpage

\end{document}